\documentclass[aps,floats,epsf,twocolumn]{revtex4}
 \usepackage{graphics}

\begin{document}

\title{Explanation for the isotropy of the Dirac cone in graphene}
\author{Igor F. Herbut}

\affiliation{Department of Physics, Simon Fraser University,
 Burnaby, British Columbia, Canada V5A 1S6}

\begin{abstract} It is shown that in the absence of spontaneous symmetry breaking the Dirac cones in the system of interacting electrons on honeycomb lattice are isotropic at low energies. The effect is due to the $Z_3$ subgroup of the $D_3$ symmetry group of the dispersion relation of Dirac quasiparticles. Consequences of the violations of the $Z_3$ or the sublattice symmetry are discussed.

\end{abstract}
\maketitle

\vspace{10pt}

   The crucial property of the single layer of graphite, i. e. graphene, is that it has well-defined, Dirac-type,  massless quasiparticles in its low-energy spectrum. It is commonly assumed that the Dirac cone is isotropic,
   with the Fermi velocity independent of direction. This is, for example, what one straightforwardly finds in simple tight-binding model of non-interacting electrons  \cite{wallace}, which is used as an input in the theories that include the Coulomb interaction. Since all the components of the electron-electron interaction, including the Coulomb's unscreened $\sim 1/r$ tail, are technically irrelevant perturbations at the non-interacting fixed point \cite{herbut}, it seems natural then to assume that the Dirac cone stays isotropic in presence of interactions as well.

         And indeed, the above expectation seems to be fulfilled in experiment \cite{lanzara}. The above simple reasoning for the isotropy of the Dirac cone, however, is faulty. The irrelevance of interactions is only a long distance property, and it is not obvious why the lattice effects, which need to be inserted by hand into the low energy theory, and would certainly renormalize the Fermi velocity, should not also break the cone's isotropy. The honeycomb lattice, after all, has only $C_{6v}$ symmetry. Clearly, if an anisotropic cone would be assumed in the low-energy theory, for example, the interactions, being irrelevant, would not alter that hypothetical feature either.

The issue is further highlighted by a recent study \cite{giuliani} in which the effects of weak interactions in a Hubbard-type model on honeycomb lattice are determined in a more rigorous form as a convergent series. The authors conclude that whereas the Fermi velocity does become renormalized by the short-range interactions, it does so in an isotropic fashion. This result comes out in a rather non-trivial analysis of the terms in the perturbation series. The utter simplicity of the conclusion, and its obvious relevance for the physics of graphene, however, beg for an elementary  explanation. Such an explanation is offered here by showing that the single-particle excitation spectrum at low energy has the exact $D_3$ symmetry, which suffices for the Dirac cone to be isotropic asymptotically. The anisotropy may be induced by the reduction in Hamiltonian's symmetry. The argument is based only on the existence of the well-defined quasiparticle excitations around Dirac points, and on the absence of the spontaneous symmetry breaking by the ground state.

   Let as assume that there exists a Dirac point in the Brillouin zone at a wavevector $\vec{K}$ near which quasiparticles are well defined in the sense of Landau. $C_{6v}$ symmetry of the {\it interacting} Hamiltonian on the honeycomb lattice implies then that the quasiparticle spectrum $e(\vec{k})$ is such that
   \begin{equation}
   e(\vec{k}+ \vec{R})= e(\vec{k}),
   \end{equation}
   and
   \begin{equation}
   e(U_\phi \vec{k})= e(\vec{k}),
   \end{equation}
   where $\vec{R}$ is any of the two reciprocal lattice vectors of the triangular Bravais lattice, and $U_\phi $ is a rotation by the integer multiple of $\pi/3$. (We will subscribe here to the notation in \cite{semenoff}.) This in particular dictates that there must be six Dirac points, related to each other by the above discrete rotations.

     Consider  a point $\vec{k}_1$ in the vicinity of the chosen Dirac point at $\vec{K}$. There are two other symmetry-related points near the same Dirac point:
     \begin{equation}
     \vec{k}_2 = U_{2\pi/3} \vec{k}_1 + \vec{R}_1,
     \end{equation}
     \begin{equation}
     \vec{k}_3 = U_{4\pi/3} \vec{k}_1 + \vec{R}_2,
     \end{equation}
   so that obviously $e(\vec{k}_1)=e(\vec{k}_2)= e(\vec{k}_3)$. These three points are related to each other by the rotations by $2\pi/3$ {\it around the Dirac point}. The Dirac cone therefore inherits the $Z_3$ subgroup of the lattice symmetry. Furthermore, since for any wavevector $\vec{k}$ we also have the reflection symmetries,
   \begin{equation}
   e(k_x, k_y) = e(-k_x,k_y)= e(k_x,-k_y),
   \end{equation}
  it follows that, in the reference frame in which the x-axis is chosen along one of the reciprocal vectors \cite{semenoff}, one of these applies also to the vicinity of the Dirac point. More precisely, if we define the momenta relative to the Dirac point as $q_x = k_x-K_x$ and $q_y = k_y - K_y$, then the reflection symmetry together with Eq. (1) implies
  \begin{equation}
  e(q_x + K_x, q_y + K_y ) =  e(-q_x + K_x , q_y + K_y).
  \end{equation}
  The full symmetry of the Dirac cone is therefore
  \begin{equation}
  D_3= Z_2 (reflection)  \times  Z_3 (rotations),
  \end{equation}
  i. e. the symmetry of the equilateral triangle.

It is convenient to define the complex wavevector measured from the Dirac point:
   \begin{equation}
q= q_x + i q_y,
\end{equation}
  \begin{equation}
\bar{q}= q_x  - i q_y,
\end{equation}
and place the origin of the coordinate frame in the reciprocal space at the Dirac point, so that $e(\vec{q}+ \vec{K})\rightarrow e(\vec{q})$.

We will assume the particle-hole symmetry of the spectrum.
In graphene, the particle-hole symmetry is weakly violated by the very small next-nearest-neighbor hopping.  At sufficiently low-energy, however, strictly speaking the particle-hole symmetry always emerges, after the chemical potential has been tuned to the Dirac point. Since in this note we are interested only in the asymptotic form of the energy-momentum relation, possible lack of particle-hole symmetry at higher energies does not affect our discussion in an essential way.

Particle-hole symmetry of the Dirac spectrum implies that one should consider the square of the energy as a power series
\begin{eqnarray}
e^2 (q,\bar{q}) = a_1 q +  b_1 \bar{q} + a_2 q^2 + b_2 q \bar{q} + c_2 \bar{q}^2 +  \\ \nonumber
a_3 q^3 + b_3 q^2 \bar{q} + c_3 q \bar{q}^2 + d_3 \bar{q}^3 + h.o.t. ,
\end{eqnarray}
where $x_i$, $x=a,b,c,d$ are constant coefficients.  The $Z_3$ symmetry of the spectrum implies that
\begin{equation}
e( e^{i n 2\pi/3}  q, e^{-in 2\pi/3} \bar{q}) = e (q,\bar{q}),
\end{equation}
for $n=1,2$, so that in the above expansion
\begin{equation}
a_1 = b_1 = a_2 = c_2 = b_3 = c_3 = 0.
\end{equation}
Together with the reflection part of $D_3$,  which demands that
\begin{equation}
e (q,\bar{q})= e (-\bar{q}, -q),
\end{equation}
this implies that
\begin{equation}
e^2 (q,\bar{q}) =  b_2 ( q\bar{q}) + a_3 (q^3 - \bar{q}^3) + c_4 (q \bar{q})^2 + O(q^4 \bar{q}, q \bar{q}^4),
\end{equation}
where we included the single allowed quartic term as well.  The reality of the expression requires the coefficients $b_2$ and $c_4$ to be real, and $a_3$ to be purely imaginary.

To the leading order the spectrum near the Dirac point is therefore isotropic, and one can identify the coefficient $b_2$ as the (squared) Fermi velocity.

   It is also interesting to consider the situation where the Hamiltonian is deformed so that the spectrum near the Dirac point remains only $Z_2$ (reflection) symmetric. This, for example, would arise if the nearest-neighbor hopping in one direction is different from the other two in the tight-binding model \cite{kohmoto}. The energy then has the form
   \begin{equation}
   e^2 (q,\bar{q}) = a_1 (q-\bar{q}) + a_2 (q^2 + \bar{q}^2) + b_2 q \bar{q} + h. o. t.
   \end{equation}
   so that, besides the shift of the location of the Dirac point (due to a finite coefficient $a_1$), the effect is the anisotropy of the Dirac cone (due to a finite coefficient $a_2$).

   Finally, breaking of the sublattice or the time-reversal symmetry \cite{herbut} would only result in the addition of a finite term to the final form of the Taylor expansion of the energy in Eq. (14), since the group $D_3$ remains the symmetry of the energy dispersion in the presence of the mass-gap.

   To conclude, whenever the Dirac quasiparticles appear in the spectrum of the interacting Hamiltonian that respects the symmetry of the honeycomb lattice, they inherit the $D_3$ subgroup of the lattice symmetry group. The $Z_3$ subgroup of the $D_3$, when not broken, then forces the Fermi velocity to be isotropic. The anisotropy appears only in the sub-leading term in the dispersion relation.

   This work has been supported by the NSERC of Canada. The author is grateful to A. Giuliani and V. Mastropietro for stimulating  correspondence.

\end{document}